\documentclass[runningheads]{llncs}
\usepackage{graphicx,color,diagbox,makecell,multirow,float,booktabs,bbding,subfigure}
\usepackage{url}
\usepackage[linesnumbered,ruled,lined,boxed]{algorithm2e}
\usepackage{algorithmic}
\begin{document}
%

\title{Flickr30K-CFQ: A Compact and Fragmented Query Dataset for Text-image Retrieval}  
\titlerunning{A Compact and Fragmented Query Dataset for Text-image Retrieval}

\author{Haoyu Liu\inst{1\star} \and
Yaoxian Song\inst{2}\thanks{Equal contribution} \and
Xuwu Wang\inst{3} \and
Xiangru Zhu\inst{3} \and
Zhixu Li\inst{3}\textsuperscript{\Envelope} \\
Wei Song\inst{1}\textsuperscript{\Envelope} \and
Tiefeng Li\inst{2}}


\institute{Research Center for Intelligent Robotics, Zhejiang Lab, China \\ \email{hyuliu20@gmail.com}, \email{weisong@zhejianglab.com}\and Department of Engineering Mechanics, Center for X-Mechanics,\\ Zhejiang University, China \\ \email{\{songyaoxian,litiefeng\}@zju.edu.cn}\and Shanghai Key Laboratory of Data Science, \\School of Computer Science, Fudan University, China \\
\email{xuwu\_wang@163.com}, \email{\{zhixuli,xrzhu19\}@fudan.edu.cn} 
}

\maketitle              
\begin{abstract}
With the explosive growth of multi-modal information on the Internet, unimodal search cannot satisfy the requirement of Internet applications. Text-image retrieval research is needed to realize high-quality and efficient retrieval between different modalities. Existing text-image retrieval research is mostly based on general vision-language datasets (e.g. MS-COCO, Flickr30K), in which the query utterance is rigid and unnatural (i.e. verbosity and formality). To overcome the shortcoming, we construct a new \textbf{C}ompact and \textbf{F}ragmented \textbf{Q}uery challenge dataset (named \textbf{Flickr30K-CFQ}) to model text-image retrieval task considering multiple query content and style, including compact and fine-grained entity-relation corpus. We propose a novel LLM-based Query-enhanced method using prompt engineering based on LLM. Experiments show that our proposed Flickr30-CFQ reveals the insufficiency of existing vision-language datasets in realistic text-image tasks. Our LLM-based Query-enhanced method applied on different existing text-image retrieval models improves query understanding performance both on public dataset and our challenge set Flickr30-CFQ with over $\mathbf{0.9\%}$ and $\mathbf{2.4\%}$ respectively. Our project can be available anonymously in \url{https://sites.google.com/view/Flickr30K-cfq}.





\end{abstract}
\section{Introduction}

Text-image retrieval refers to the process of retrieving information across different data modalities (e.g. text-image, video-text, audio-text), which has been applied in various fields (e.g. multimedia information retrieval~\cite{cheng2020robust}, recommendation systems~\cite{chen2021cmbf}, smart assistants and human-computer interaction~\cite{zhen2023human}, etc.). Technically, it involves searching for relevant content in one modality based on a query from a different modality by extracting discriminative features and summarizing information from multiple modalities~\cite{wang2023balance}. Traditional retrieval methods focus on separate feature extraction for different modalities and specific matching or similarity measures~\cite{wangAdversarialCrossmodalRetrieval2017,weiCrossmodalRetrievalCNN2016}. With the explosive growth of large language models (LLMs), more and more work is considering to use pre-trained models to learn robust representation and prediction models~\cite{lu2019vilbert,kim2021vilt}.

\begin{figure}[t]
  \centering
  \includegraphics[width=0.85\linewidth]{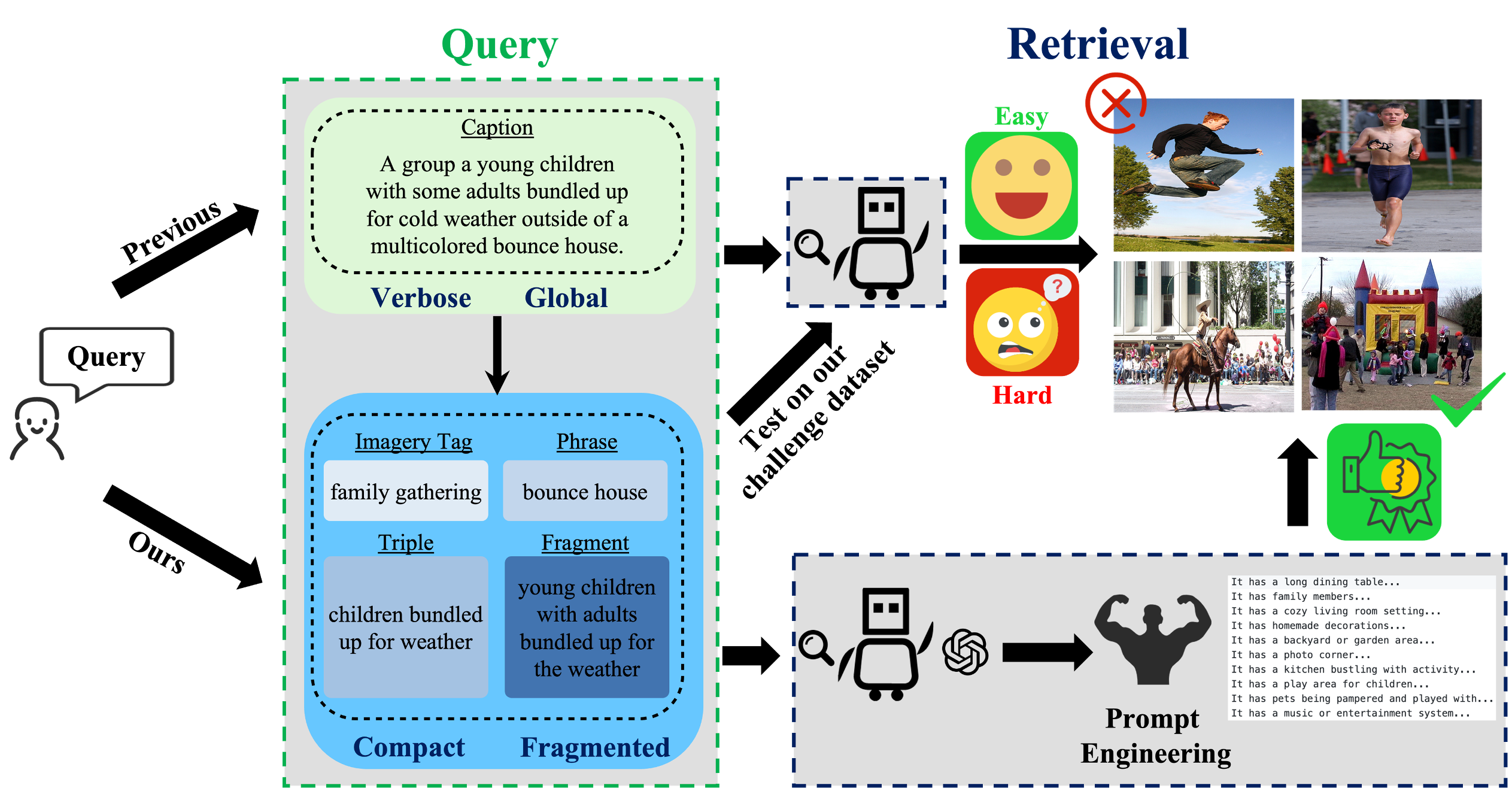}
  \caption{The overview of text-image models. (1) \textbf{Previous:} The query in existing datasets is verbose and global caption and the retrieval models unitize the query directly. (2) \textbf{Ours:} Our dataset contains four-level granularities corpus and proposed model uses LLMs to enhance the compact and fragmented query for subsequent retrieval.}
  \label{fig:overview}
\end{figure}

Existing dataset and benchmarks~\cite{youngImageDescriptionsVisual2014a,linMicrosoftCOCOCommon2015a} for text-image retrieval are still difficult to meet real-world task requirements. Cross-modal alignment is one of the core problems needed to solve in text-image retrieval, for which existing work usually adopts general vision-language data, such as MS-COCO~\cite{linMicrosoftCOCOCommon2015a} and Flickr30K~\cite{youngImageDescriptionsVisual2014a}, in which the text is verbose and formal leading to decreasing performance on retrieval scenarios. From the granularity, the query in an existing dataset usually provides a global or coarse-grained description of the image, while a user prefers to use compact words or fragments to search for information on a practical scenario. Secondly, from the length of a query, the content of a query may be abundant. For instance, as shown in Fig.~\ref{fig:overview}, query depicts ``\textit{A group a young children with some adults bundled up for cold weather outside of a multicolored bounce house.}" in Flickr30K, in which the expression is in written form and abundant copulas appear frequently. In contract, people usually use ``family gathering, bounce house, children bundled up for weather, etc." to inquire the target images in oral speaking free from grammar and voice restrictions. For length of sentence, the average number of tokens in Flickr30K~\cite{youngImageDescriptionsVisual2014a}, MS-COCO~\cite{linMicrosoftCOCOCommon2015a}, and LAIT~\cite{qiImageBERTCrossmodalPretraining2020a} are $13.4$, $10.4$ and $13.4$ respectively. By contrast, statistics on ORCAS~\cite{craswellORCAS18Million2020}, a search log based on real-world scenarios, shows that each query contains $3.2$ tokens on average.

To overcome the limitation of query form in real-world scenario, we propose a novel comprehensive vision-language dataset, named Flickr30K-CFQ, by extending typical vision-language dataset~\cite{plummerFlickr30kEntitiesCollecting2016a} with compact and fragmented corpus for the natural query retrieval. We consider two challenges (i.e. \textit{oral-compact} expression and \textit{local-fragmented} query) in the pragmatic dataset for text-image retrieval problem. For the local query we introduce \textit{\textbf{triple}} (entity \& relation), and \textit{\textbf{fragments}} (multiple triples) corpus. For oral-compact expression, \textit{\textbf{imagery tag}} (abstract) and \textit{\textbf{phrase}} are given. Specifically, imagery tag is produced by large multi-modal model LLaVA~\cite{liu2023visual}, which generates a series of abstract imagery descriptions. We incorporate the manually annotated noun phrases from the Flickr30K Entities~\cite{plummerFlickr30kEntitiesCollecting2016a} dataset as a specific query type. We utilize the StanfordNLP OPENIE component~\cite{manningStanfordCoreNLPNatural2014} to extract Subject-Predicate-Object (SPO) triples, 
which are then employed to create fragments. Multiple triples are fused to generate a \textit{fragment} by a fine-tuned Google T5~\cite{raffelExploringLimitsTransfer2020a}. In all, we provide four-level granularities query corpus as shown in Fig.~\ref{fig:overview}, which ranges from abstract to concrete and from global to local query.

For modeling text-image retrieval, existing methods have not considered the aforementioned challenges.
Existing research on text-image retrieval can be roughly divided into two types according to whether to utilize pre-trained models~\cite{rao2022does}. Methods not utilizing pre-trained models usually focus on improving modal fusion\cite{faghri2017vse++,wang2019camp,lee2018stacked} and similarity modeling process~\cite{diao2021similarity,li2019visual,wu2019learning,zhang2020context}. They use limited learning parameters to obtain satisfactory performance in specific domain tasks, while generalizing poorly in open-world applications. For methods using vision-language pre-trained models (VLP), they utilize multi-modal semantic priority knowledge in pre-trained models to model multi-modal alignment in text-image retrieval tasks~\cite{li2021align,jia2021scaling,lüddeckeImageSegmentationUsing2022,wang2022image}. They have advantages in both performance and generalization compared to traditional train-from-scratch methods. 
Although existing text-image retrieval research has achieved impressive performance~\cite{luViLBERTPretrainingTaskAgnostic2019,liVisualBERTSimplePerformant2019a,li2020oscar,zeng2021multi}, we find that it still faces the problem of unnatural textual query on real-world scenarios, which deriving requirements for robust query understanding about compact or fragmented text. Therefore, to improve this natural query retrieval performance, we propose a novel query-enhanced retrieval framework using LLM-based prompt engineering, shown in Fig.\ref{fig:overview}. Compact or fragmented queries are extended into a batch of comprehensive queries using prompt engineering, used to model cross-modal alignment instead of solo query input. Open-source and commercial LLMs are used respectively during the modeling process to evaluate the effectiveness of our proposed method.

Our contributions can be summarized as follows:
\begin{enumerate}
    \item We introduce a new Compact and Fragmented Query dataset to the text-image retrieval community, named \textbf{Flickr30K-CFQ}, which is used to model natural text-image retrieval in real-world scenarios. It could make up for the deficiency of the existing text-image retrieval vision-language dataset in verbosity and formality corpus.
    \item An LLM-based Query-enhanced text-image retrieval method for natural query scenarios is proposed. It adopts prompt engineering based on LLMs to augment compact or abstract input query. By multi-turn voting mechanism, our method obtains stable augmentation performance to improve the robustness.
    \item Experiments on query-enhanced variants based on our proposed method using open-sourced and commercial LLMs show the effectiveness of our method and achieve obvious improvement with over $0.9\%$ and $2.4\%$ respectively on public dataset and our challenge set Flickr30-CFQ. Comparisons between existing dataset and our Flick30K-CFQ indicate that our proposed dataset reveal the insufficiency existing dataset for text-image retrieval research and the necessity of our Flick30K-CFQ.
\end{enumerate}

\section{Related Work}
In this paper, we propose a novel comprehensive dataset Flickr30-CFQ and query-enhanced text-image retrieval model. We review the existing work from dataset construction and data augmented text-image retrieval models.

\subsection{Datasets for Text-image Retrieval}
For text-image retrieval, various multi-modal datasets have been developed to train models and evaluate retrieval techniques~\cite{peng2017overview}, which can be categorized into extended-based dataset and original dataset. For the former, prominent among these are the Flickr30K~\cite{youngImageDescriptionsVisual2014a} and MS-COCO~\cite{linMicrosoftCOCOCommon2015a,chenMicrosoftCOCOCaptions2015} datasets, widely regarded as benchmarks. Additionally, the datasets of NUS-WIDE~\cite{chua2009nus} and Wikipedia~\cite{rasiwasia2010new} apply to specific research scenarios.
The Flickr30K dataset comprises $31,783$ images sourced from Flickr, each supplemented with five descriptive captions generated through crowdsourcing. For another, MS-COCO dataset is designed to emphasize daily life scenes. It includes $123,287$ images, each featuring at least five human-generated descriptions, aiming to capture diverse real-world contexts. Diverging from the approach of Flickr30K and MS-COCO, the NUS-WIDE~\cite{chuaNUSWIDERealworldWeb2009} dataset encompasses $269,648$ images from Flickr, annotated with around $5,018$ unique single-word tags manually.
For the latter, large-scale text-image datasets collected from scratch has been attempted recently. One of these is the LAIT~\cite{qiImageBERTCrossmodalPretraining2020a}, which employs heuristic rules to filter Internet-scale data, pairing images with user-defined HTML metadata as captions. LAIT uses a modest amount of supervised data to ensure the semantic alignment between images and texts.

When examining the aforementioned datasets, we find that most queries offer a global description of each paired image or an artificial written expression, which is unnatural and rigid in the varied and flexible real-world scenarios compared to the query by human. To address this, we consider to collect compact, fragmented and fine-grained descriptions adapting to the human query style, which is organized as dataset Flickr30K-CFQ. It could not only facilitates the training of more contextually relevant text-image retrieval models but also serves as a new and challenging benchmark.

\subsection{Data Augmented Text-image Retrieval Model}
One of the conventional data augmentation techniques focuses on generating additional training data to improve model performance in text-image retrieval in offline mode. They augment hard negative samples to achieve better performance by contrastive learning~\cite{xia2021progcl}.
Different strategies are employed for hard negative samples generation. Visual-Semantic Embedding (VSE) model~\cite{kiros2014unifying} utilizes random sampling to select hard negatives.
Based on it, the most challenging samples within a batch are considered in VSE++~\cite{faghri2017vse++} effectively to extend the selection space to encompass the entire dataset. After that, 
The Adaptive Object Query (AOQ) model~\cite{chen2020adaptive} is used to refine the challenging sampling policy by selecting from all training data directly instead of a batch of data within all training data, leveraging pre-trained models. Furthermore, TAGS-DC~\cite{fan2021negative} proposes a counterfactual method, in which keywords in a positive sentence are modified to derive hard negatives. 
On the other hand, the external knowledge guided method is explored to enhance the query in online mode. Cakp~\cite{li2022supporting} enhances the semantics of the initial query by integrating an ontology knowledge graph retrieving information. MKVSE~\cite{feng2023mkvse} employs a Multi-Modal Knowledge Graph (MMKG) to construct implicit relationships between images and texts, mainly when the image encompasses information not explicitly described in the accompanying text.

\begin{figure}[t]
  \centering
  \includegraphics[width=0.8\linewidth]{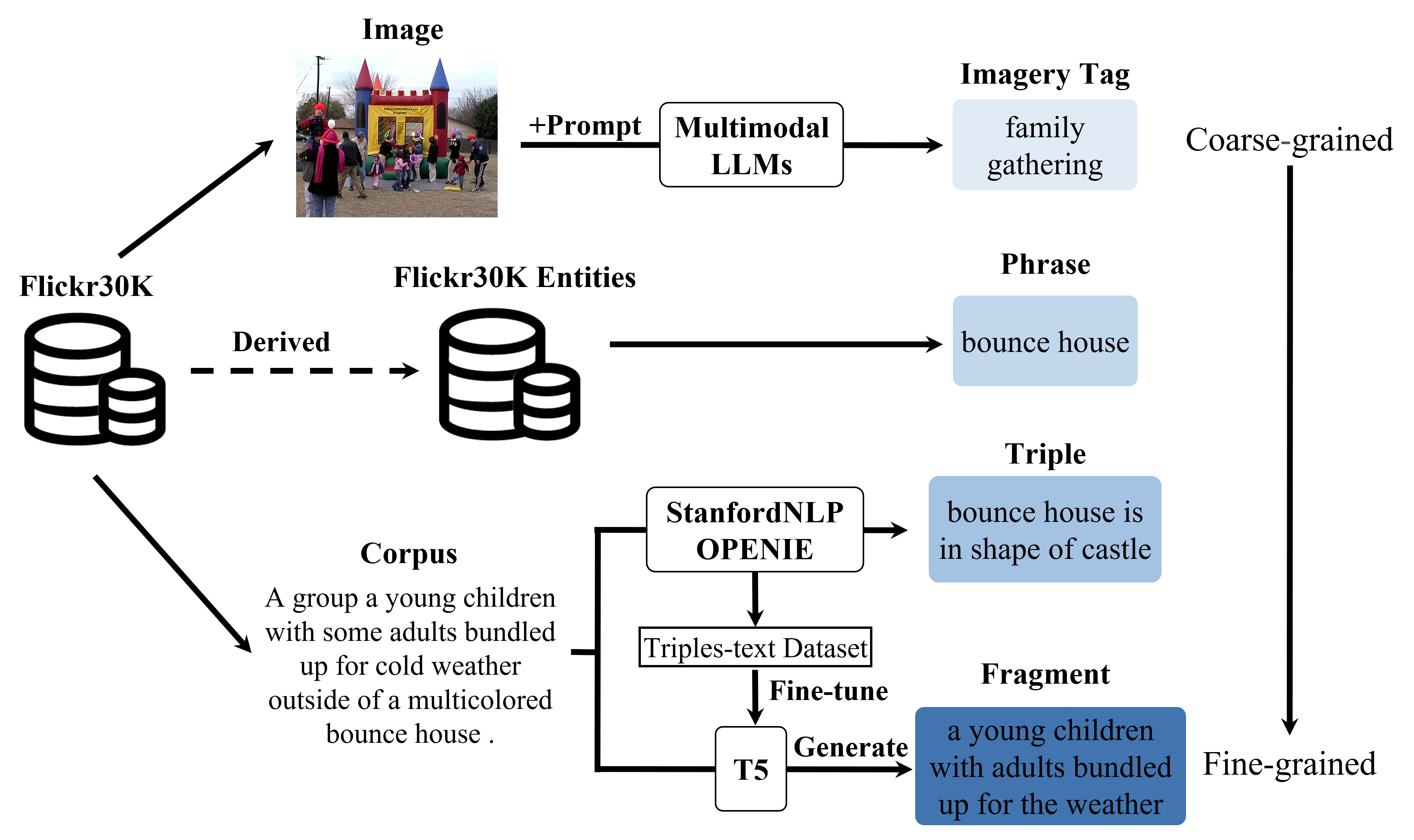}
  \caption{The Construction of Flickr30K-CFQ. Our dataset provides four-level granularities query corpus: 1). \textbf{Imagery Tag} (abstract) is annotated by multi-modal LLM. 2). \textbf{Phrase} is inherited from Flickr30K Entities 3). \textbf{Triple} (entity \& relation) is extracted from corpus. 4).\textbf{Fragment} (multiple triples) is generated by the fine-tuned T5 based on multiple SPO.}
  \label{fig:process}
\end{figure}

For offline-based methods, they enhance the training data by additional hard negative samples or counterfactual data augmentation~\cite{yang-etal-2023-distribution}, constrained by the negative semantic text generation. It could neither realize data enhancement dynamically. For online-based methods, explicit and structured knowledge from external knowledge base are adopted, which relies on high-quality domain-specific knowledge or commonsense knowledge. This results in poor generalization in different scenarios. To overcome the above limitations, we attempt to use LLMs to design a query-enhanced text-image retrieval model, which leverages implicit knowledge in pre-trained models~\cite{brownLanguageModelsAre2020,VicunaOpenSourceChatbot} to generate an amount of unstructured related queries for the retrieval task.

\begin{table}[t]
\caption{Statistics of Flickr30K-CFQ. Our Flickr30K-CFQ fills the gap of oral-compact expression and local-fragmented query}\label{tab:statistics}
\centering
\begin{tabular}{@{}cl|ccccc@{}}
\toprule
Dataset              & Image  & Caption & Imagery Tag & Phrase  & Triple  & Fragment \\ \midrule
Flickr30K            & 31,783 & 158,915 & \XSolidBrush           & \XSolidBrush       & \XSolidBrush       & \XSolidBrush        \\ \midrule
Flickr30K-CFQ (\textbf{ours}) & 31,783 & 158,915 & 305         & 111,240 & 133.540 & 139,607  \\ \bottomrule
\end{tabular}
\end{table}

\section{Dataset Construction}
We introduce a new \textbf{C}ompact and \textbf{F}ragmented \textbf{Q}uery challenging dataset (named \textbf{Flickr30K-CFQ}) based on Flickr30K Entities~\cite{plummerFlickr30kEntitiesCollecting2016a}. Our Flickr30K-CFQ encompasses three key facets: (1) \textbf{Fundamental Concept of Flickr30K-CFQ}, (2) \textbf{Construction Pipeline}, (3) \textbf{Statistical Analysis and Comparisons}.

\subsection{Fundamental Concept of Flickr30K-CFQ}\label{Flickr30K-CFQ}
Existing queries in text-image retrieval datasets are rigid and global-depiction. To address this gap, we collect local and fine-grained queries: (\textbf{Triple} and \textbf{Fragment}), and oral and compact expressions: (\textbf{Imagery Tag} and \textbf{Phrase}), which is named \textbf{C}ompact and \textbf{F}ragmented \textbf{Q}uery challenge dataset (\textbf{Flickr30K-CFQ}). The characteristics of these query corpus are defined as below:

\begin{itemize}
    \item[\textbullet\quad] \textbf{Caption:} Our dataset includes the caption from Flickr30K Entities~\cite{plummerFlickr30kEntitiesCollecting2016a} (inherited from Flickr30K~\cite{youngImageDescriptionsVisual2014a}). The caption describes an image in a global scope, and its text expression is unnatural and redundant.
    \item[\textbullet\quad] \textbf{Imagery Tag:} The imagery tag is an abstract and compact short text of the image. Users employ tags such as ``pleasant afternoon" and ``family gathering" to retrieve corresponding images.
    \item[\textbullet\quad] \textbf{Phrase:} The phrase is also from Flickr30K Entities~\cite{plummerFlickr30kEntitiesCollecting2016a}. It is a noun phrase, describing the concrete entity about the image.
    \item[\textbullet\quad] \textbf{Triple:} Triple is SPO triple describing corpus about relationships between a part of an image or contains entities. It provides more fine-grained relationship information for query instead of entity only in phrase.
    \item[\textbullet\quad] \textbf{Fragment:} Fragment is composed of multiple triples with more various fine-grained description about retrieved image.
\end{itemize}

\subsection{Construction Pipeline}\label{construction}
As shown in Fig.~\ref{fig:process}, our Flickr30K-CFQ introduce four-granularity corpus. For \textbf{Imagery tags}, we design multiple prompts for multi-modal large language model~\cite{liu2023visual} to generate abstract and compact tag related to the target image from the Flickr30K~\cite{youngImageDescriptionsVisual2014a}. Next, for \textbf{Phrase}, we obtain the corpus from existing dataset Flickr30K Entities~\cite{plummerFlickr30kEntitiesCollecting2016a}, which is about entity-level description and used in visual language grounding task originally~\cite{li2022grounded}. In the following step, we firstly fine-tune a T5 model using self-owned triples-text dataset~\cite{raffelExploringLimitsTransfer2020a}, which is then used to extract various triples. Individual triple is adopted as \textbf{Triple}. Multiple triples are combined as \textbf{Fragment}. Specifically, compared to \textbf{Caption} in original Flickr30K, our method provide rich corpus ranging from abstract to concrete and from global to local query.

\subsection{Statistical Analysis and Comparison} \label{sec:statistics}
The comparison of statistical characteristics of between our Flickr30K-CFQ and original Flickr30K are denoted in Table~\ref{tab:statistics}. Apart from original $148,915$ \textbf{Caption} from Flickr30K, our Flickr30K-CFQ introduce additional $305$ \textbf{Imagery Tag}, $111,240$ \textbf{Phrase}, $133,540$ \textbf{Triple}, and $139,607$ \textbf{Fragment}. In all, our new Flickr30K-CFQ expands over \textbf{three times} corpus compared original Flickr30K with five different granularities.

\begin{figure}[t]
  \centering
  \includegraphics[width=0.9\linewidth]{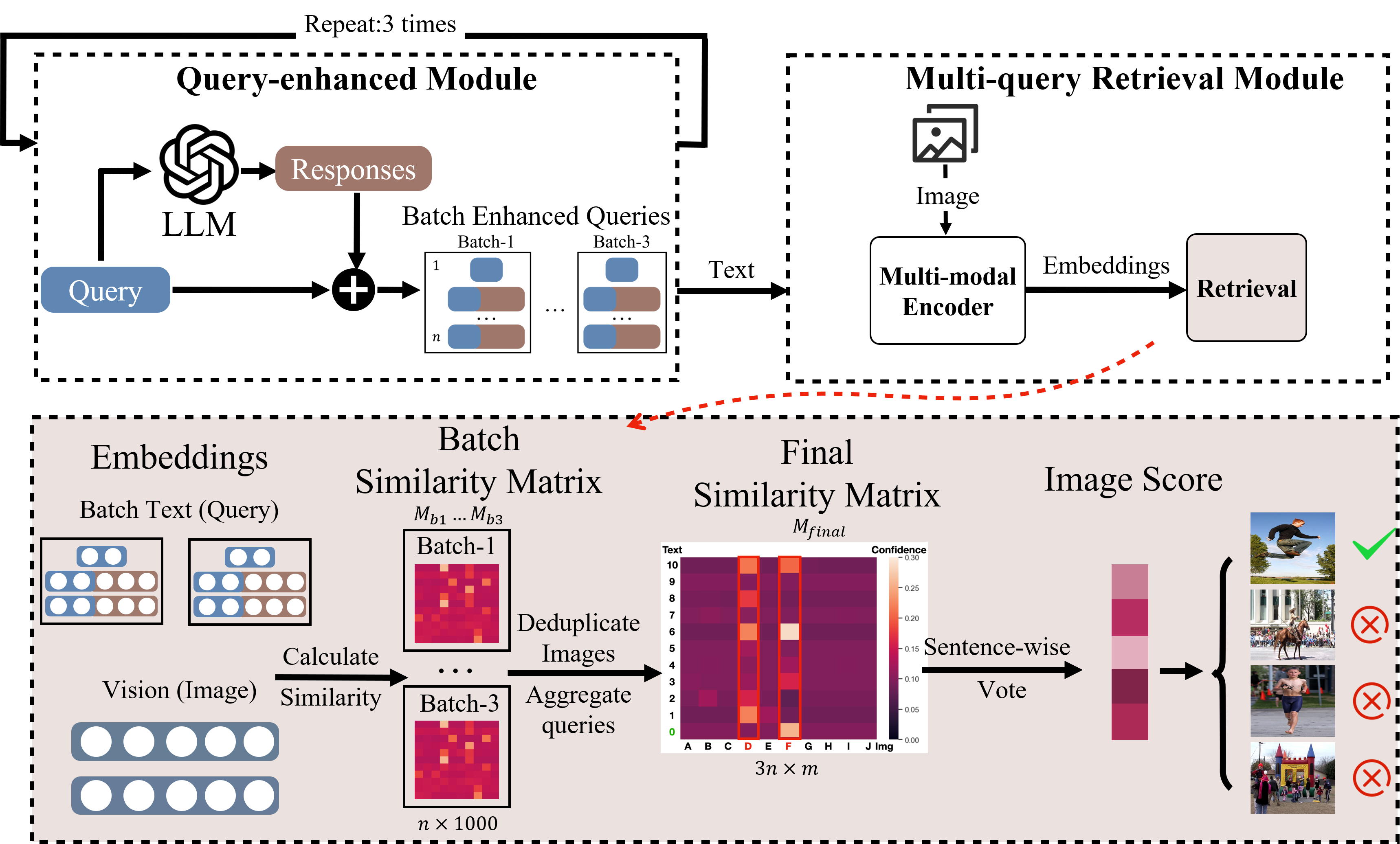}
  \caption{LLM-based Query-enhanced method including two modules. The first is \textbf{Query-enhanced Module}, which is used to expand the initial query to a query batch. The second is \textbf{Multi-query Retrieval Module}, in which two-stage similarity are calculated to obtain better retrieved results.}
  \label{fig:model}
\end{figure}

\section{LLM-based Query-enhanced Method}
We propose a LLM-based Query-enhanced method, augmenting the potential semantic information for the input query by prompt engineering. The overview of our method is shown in Fig.~\ref{fig:model}. It consists of two modules, in which the first is \textbf{Query-enhanced Module} to generate various corpus related to initial query. The second is \textbf{Multi-query Retrieval Module} to predict retrieved images.

\subsection{Query-enhanced Module}\label{sec:query_enhanced}
As shown in the top-left of Fig.~\ref{fig:model}, large language models (LLMs) are employed to stretch the original input query based on the pre-trained knowledge in LLMs. Specifically, multiple handcrafted prompts are designed to induce the LLM to generate sentences by prompt learning, which are both related to the input query and target retrieved image. The expanded sentences are concatenated with the initial input query as the whole query-enhanced input to the text encoder, shown in the top-right of Fig.~\ref{fig:model}. Furthermore, due to the randomness in LLM generation, we repeat the augmentation operation multiple times to obtain reliable batches of enhanced sentences~\footnote{$3$ times in this work.}.

\subsection{Multi-query Retrieval Module}
Using the enhanced queries in Sec.~\ref{sec:query_enhanced} and paired candidate images, we train our query-enhanced retrieval models based on \textbf{Multi-query Retrieval Module}. It contains multi-modal feature (Encoder) extraction and retrieval, as shown in the top-right of Fig.~\ref{fig:model}. A multi-modal pre-trained model~\cite{xuGroupViTSemanticSegmentation2022,lüddeckeImageSegmentationUsing2022,jiaScalingVisualVisionLanguage2021d,radford2021learning} is selected to encoder textual and visual features, which are used to calculate cosine similarity pairwise. 

To obtain more reliable retrieval shown in the bottom of Fig.~\ref{fig:model}, for each batch of expanded sentences, we firstly obtain a series of similarity matrix $M_{b1}...M_{b3} \in \mathbf{R}^{n \times 1000}$ ($1000$ is the initial number of candidate images; we obtain $3$ batch of the enhanced sentence in Sec.\ref{sec:query_enhanced}; $n$ is the number of the sentence in a batch) and filter out a new candidate image set ($1,000 \rightarrow 15$) by \textit{Top@K} for each sentence in the batch ($n \times 15$ images). After that, we remove duplicates and aggregate all new sets from each batch as the final candidate image set containing $m$ images over $3$ batches. Then, images in the final set and $n \times 3$ expanded sentences are used to calculate cosine similarity again to obtain similarity matrix $M_{final} \in \mathbf{R}^{3n \times m}$. Finally, \textit{Top@K} votes are made in the sentence wise to get the final retrieved results.

\section{Experiment}
We evaluate our proposed Flickr30K-CFQ and query-enhanced method respectively. Firstly, experiments tested on Flickr30K-CFQ by SOTA method with query enhancement are given to evaluate the necessity of the proposed dataset and the effectiveness of query-enhanced method comprehensively. Secondly, the generalization of our query-enhanced method using both commercial and open-sourced large language models is analyzed. Thirdly, the performance of the query-enhanced method in the public text-image dataset are introduced, which analyzes the dependency level of the dataset for our method.

\subsection{Implementation Details} \label{sec:details}
\subsubsection{Models:} We select four representative multi-modal pre-trained models:
\\ \textbf{GroupViT~\cite{xuGroupViTSemanticSegmentation2022}}, \textbf{CLIPSeg~\cite{lüddeckeImageSegmentationUsing2022}}, \textbf{ALIGN~\cite{jiaScalingVisualVisionLanguage2021d}}, \textbf{CLIP~\cite{radford2021learning}}\footnote{groupvit-gcc-yfcc, clipseg-rd64-refined, align-base, clip-vit-base-patch32 are used in our work.}.

\subsubsection{Query-enhanced Model Setting:} For LLM in our query-enhanced method, we utilize the open-source Vicuna~\cite{VicunaOpenSourceChatbot} and the commercial model GPT-3.5~\cite{brownLanguageModelsAre2020}\footnote{We use vicuna-13b-v1.1 and gpt-3.5-turbo in our work.}. Vicuna-based experimental results are given, except Table~\ref{tab:main}. Experiments are implemented on Ubuntu 20.04.6 LTS and PyTorch 1.12.1 with four NVIDIA GeForce RTX 3090 GPUs.

\subsubsection{Evaluation Data} 
To perform our zero-shot evaluation, we randomly select $100$ sentences in our Flickr30K-CFQ paired with corresponding images (approx. $500$ images), which is as the test set for our experiment.

\subsubsection{Metrics:} 
We evaluate retrieval performance using two metrics: the traditional Recall@K and our newly proposed Multi-recall@K. While the traditional Recall@K metric is typically suited for one-to-one retrieval, it shows limitations when applied to our approach. Therefore, we introduce the Multi-recall@K metric, which is designed for one-to-many retrieval. Details of this metric are provided in Algorithm~\ref{alg:metric}, with an implementation setting of K=$10$.

\begin{algorithm}[tbp]
 \small
 \caption{Multi-recall@K}
 \label{alg:metric}
 \textbf{Input}:Predict image set $P$, true image set $T$, number c\\
 \textbf{Output}:Multi-recall@K.
 \begin{algorithmic}[1] 
    \STATE Let $c=0$;
    \FOR{$image$ in $P$}
        \IF {$image$ belongs to $T$}
        \STATE $c=c+1$;
        \ENDIF
    \ENDFOR
    \STATE Multi-recall@K = $\frac{c}{\max(K, len(P))}$;
 \end{algorithmic}
\end{algorithm}

\begin{table}[t]
\caption{Comparison in Flickr30K-CFQ with five levels of granularity. Off-the-shell models have poor performance on our challenge dataset.}
\centering
\label{tab:Result_type}
\resizebox{0.95\linewidth}{!}{
\begin{tabular}{@{}cccccc@{}}
\toprule
                 & Caption (\%)                                           & Imagery Tag (\%)                                          & Phrase (\%)                                            & Triple (\%)                                            & Fragment (\%)                                               \\ \midrule
GroupViT~\cite{xuGroupViTSemanticSegmentation2022}            & 77.58                                         & 26.62                                                  & 41.53                                                  & 56.36                                                  & 58.29                                                  \\ \midrule
Enhanced w/ vote (\textbf{ours}) & \begin{tabular}[c]{@{}c@{}}80.60\\ ($\uparrow$3.02)\end{tabular} & \begin{tabular}[c]{@{}c@{}}\textbf{27.91}\\ ($\uparrow$1.29)\end{tabular} & \begin{tabular}[c]{@{}c@{}}\textbf{44.26}\\ ($\uparrow$2.73)\end{tabular} & \begin{tabular}[c]{@{}c@{}}\textbf{56.95}\\ ($\uparrow$0.59)\end{tabular} & \begin{tabular}[c]{@{}c@{}}\textbf{61.61}\\ ($\uparrow$3.32)\end{tabular} \\ \bottomrule
\end{tabular}}
\end{table}

\begin{figure}[h]
    \centering
    \subfigure[Visualization in Flickr30K-CFQ.]{
        \label{fig:cfq}
        \includegraphics[width=0.7\linewidth]{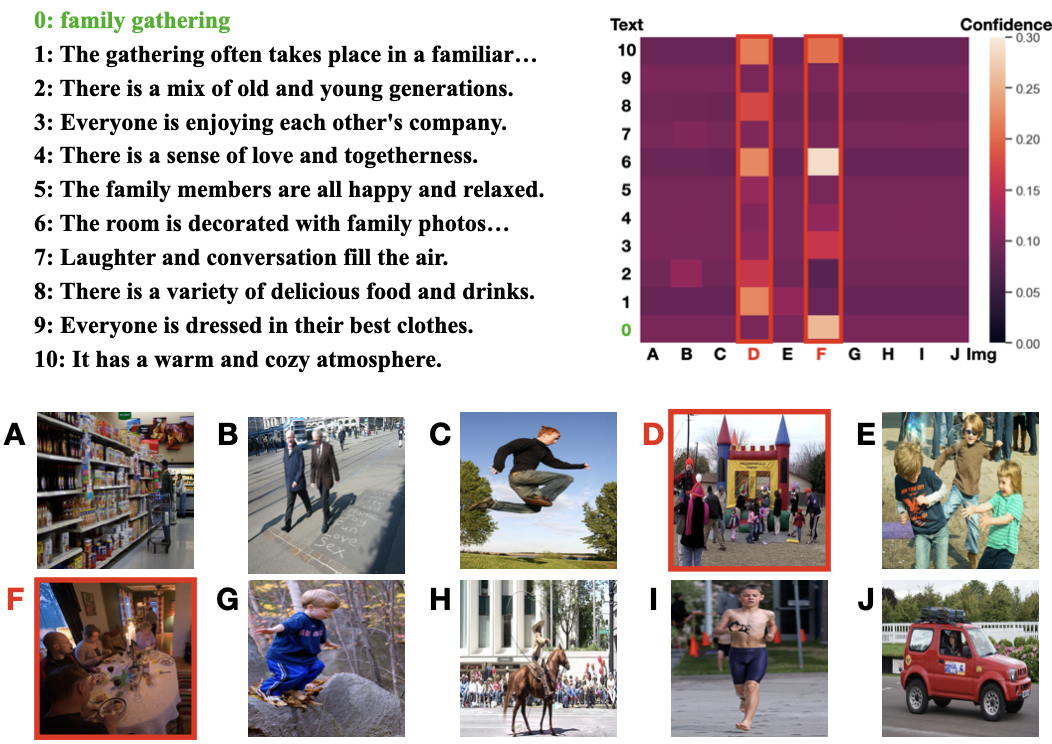} 
    }
    \subfigure[Visualization in Flickr30K.]{
        \label{fig:flickr30k}
        \includegraphics[width=0.7\linewidth]{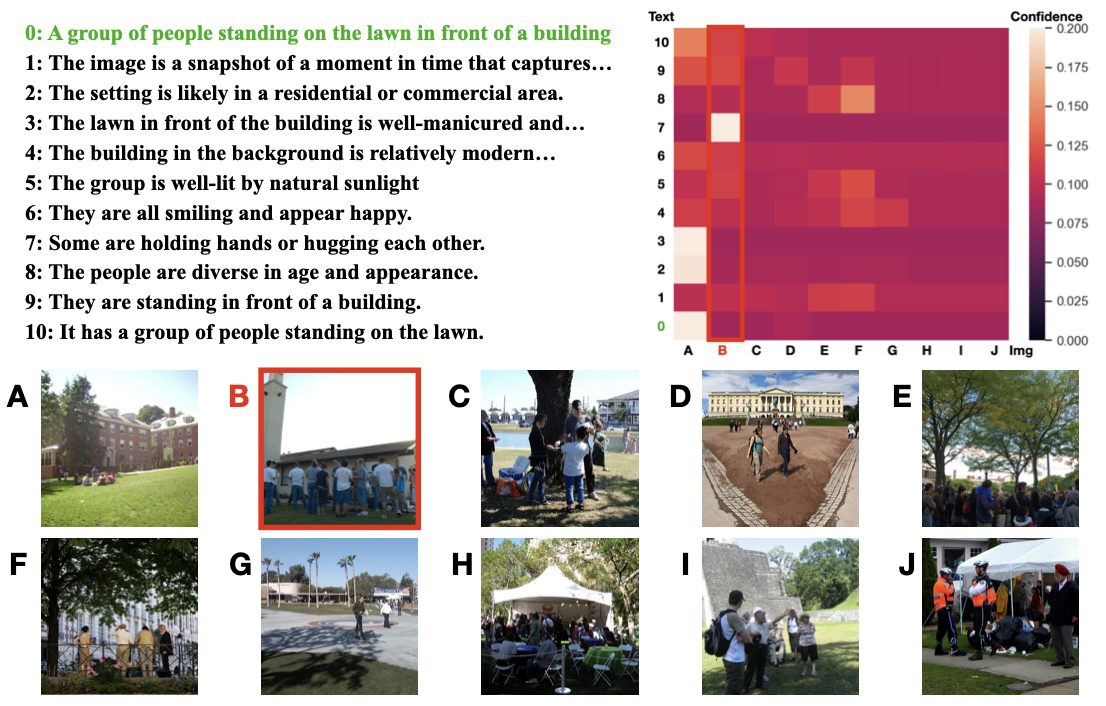} 
    }
    \DeclareGraphicsExtensions.
    \caption{Visualization of retrieval results from LLM-based Query-enhanced method.}  \label{fig:case-study}
\end{figure}

\subsection{Experimental Results} \label{sec:result}

Based on a zero-shot setting, we first compare the results on all five-level granularities query before and after enhancement. Secondly, we conduct experiments on our sub-dataset using the open-source model Vicuna~\cite{VicunaOpenSourceChatbot} and the commercial model GPT3.5~\cite{brownLanguageModelsAre2020} with four pre-trained multi-modal models. Finally, we further verify the effectiveness of the LLM-based Query-enhanced method on the benchmark dataset of text-image retrieval.

\subsubsection{Fine-grained Text-image Retrieval Evaluation}

We compare the retrieval performance before and after enhancement using five-level granularities queries as inputs. Our experiments are conducted on the Flickr30K-CFQ dataset, and we evaluate performance using the Multi-recall@10 metric for one-to-many retrieval. The results demonstrates the poor performance of off-the-shell models on retrieval tasks with compact or fragmented as queries in Flickr30K-CFQ, including imagery tags, phrases, triples and fragments. These new queries are more challenging than caption-like queries in existing benchmarks.

Current multi-modal pre-trained models perform well in text-image retrieval tasks with caption-like queries as inputs, achieving a Multi-recall@10 score of up to $78$\%. However, their performance significantly drops when utilizing four types of compact and fragmented queries from the Flickr30K-CFQ dataset, with the Multi-recall@10 score falling below $60$\% and dropping to as low as $26.62$\% when using imagery tags. The significant difference in performance indicates that the Flickr30K-CFQ dataset is more challenging compared to existing benchmarks, which are very simple for current text-image retrieval models.

\begin{table}[t]
\caption{Experiments in Flickr30K-CFQ. We compare the performance in open-source and commercial, respectively, and our method obtains SOTA in Multi-recall@10.} \label{tab:main}
\centering
\resizebox{0.9\linewidth}{!}{
\begin{tabular}{@{}ccccccccc@{}}
\toprule
                 & \multicolumn{4}{c}{Vicuna-13B}                                                                                                                                                                                                     & \multicolumn{4}{c}{GPT3.5}                                                                                                                                                                                                       \\ \midrule
                 & GroupVit                                               & CLIPSeg                                                & ALIGN                                                  & CLIP                                                    & GroupVit                                               & CLIPSeg                                                & ALIGN                                                  & CLIP                                                   \\ \midrule
Baseline         & 52.08                                                  & 66.22                                                  & 72.43                                                  & 64.31                                                   & 50.87                                                  & 65.89                                                  & 72.11                                                  & 64.84                                                  \\ \midrule
Enhanced (\textbf{ours})        & \begin{tabular}[c]{@{}c@{}}53.57\\ ($\uparrow$1.49)\end{tabular} & \begin{tabular}[c]{@{}c@{}}66.64\\ ($\uparrow$0.42)\end{tabular} & \begin{tabular}[c]{@{}c@{}}72.45\\ ($\uparrow$0.02)\end{tabular} & \begin{tabular}[c]{@{}c@{}}64.22\\ ($\downarrow$0.09)\end{tabular} & \begin{tabular}[c]{@{}c@{}}53.82\\ ($\uparrow$2.95)\end{tabular} & \begin{tabular}[c]{@{}c@{}}68.10\\ ($\uparrow$2.21)\end{tabular} & \begin{tabular}[c]{@{}c@{}}73.19\\ ($\uparrow$1.08)\end{tabular} & \begin{tabular}[c]{@{}c@{}}65.72\\ ($\uparrow$0.88)\end{tabular} \\ \midrule
Enhanced w/ vote (\textbf{ours}) & \begin{tabular}[c]{@{}c@{}}\textbf{54.28}\\ ($\uparrow$2.20)\end{tabular} & \begin{tabular}[c]{@{}c@{}}\textbf{66.84}\\ ($\uparrow$0.62)\end{tabular} & \begin{tabular}[c]{@{}c@{}}\textbf{72.72}\\ ($\uparrow$0.29)\end{tabular} & \begin{tabular}[c]{@{}c@{}}\textbf{64.71}\\ ($\uparrow$0.40)\end{tabular}  & \begin{tabular}[c]{@{}c@{}}\textbf{54.59}\\ ($\uparrow$3.72)\end{tabular} & \begin{tabular}[c]{@{}c@{}}\textbf{68.19}\\ ($\uparrow$2.30)\end{tabular} & \begin{tabular}[c]{@{}c@{}}\textbf{73.49}\\ ($\uparrow$1.38)\end{tabular} & \begin{tabular}[c]{@{}c@{}}\textbf{66.66}\\ ($\uparrow$1.82)\end{tabular} \\ \bottomrule
\end{tabular}}
\end{table}

\begin{table}[t]
\caption{Comparisons of metrics.}\label{tab:metric}
\centering
\begin{tabular}{@{}ccccc@{}}
\toprule
                & GroupVit & CLIPSeg & ALIGN  & CLIP \\ \midrule
Recall@10       & 61.76 & 74.24   & 80.67 & 71.82    \\ \midrule
Multi-recall@10 & 52.08 & 66.22   & 72.43 & 64.31    \\ \bottomrule
\end{tabular}
\end{table}

\subsubsection{Commercial vs. Open-sourced Model in Flickr30K-CFQ} To validate the effectiveness of our LLM-based Query-enhanced method, we evaluate the performance on both open-source model Vicuna-13B ~\cite{VicunaOpenSourceChatbot} and the commercial model GPT-3.5~\cite{brownLanguageModelsAre2020}.

Our proposed model achieve good performance in Table~\ref{tab:main} not only proves the efficacy and robustness of our proposed model but also validates the effectiveness of the voting mechanism. The scores of Multi-recall@10 show an average improvement of $1.12\%$ over the baseline without a voting mechanism and an improvement of $1.58\%$ after introducing the voting system. The model (with vote) achieves improvements of $0.88\%$ and \textcolor{red}{$\mathbf{2.31\%}$} on open-source and commercial models, respectively. The results also indicate that more powerful LLMs can achieve better performance with our method.

We also demonstrate the effectiveness of Multi-recall@K. On the one hand, the traditional Recall@K metric is appropriate for the one-to-one scenario, and Multi-recall@K is suitable for the one-to-many scenario, and degradation to Recall@K in the one-to-one scenario. On the other hand, our metric can also addresses the shortcomings of traditional ones and offers more rigorous criteria. According to the Table~\ref{tab:metric}, Multi-recall@K typically experiences a performance decline of over $7\%$ when in identical experimental conditions.

\subsubsection{LLM-based Query-enhanced Method in Public Benchmark} We evaluate the performance improvements in widely used benchmarks and select Flickr30K\cite{youngImageDescriptionsVisual2014a} dataset and use Recall@K(K = $1,5,10$) as metrics. 

The improvement of four models and three metrics are illustrated in Table~\ref{tab:Flickr30K}. We achieve performance improvements of $0.35$\%, \textcolor{red}{$\mathbf{0.98\%}$}, and $0.91$\% on Recall@1, Recall@5, and Recall@10. The most significant improvement was observed in the GroupVit model, with a \textcolor{red}{$\mathbf{2.14\%}$} increase in the Recall@5. The outcomes demonstrate that the LLM-based Query-enhanced method can compensate for the text's semantic information deficiency and deliver more valuable text for the pre-trained text-image retrieval models.

\begin{table}[t]
\caption{Comparisons of different models in Flickr30K. Models based on our method archive the better performance widely.}\label{tab:Flickr30K}
\centering
\begin{tabular}{@{}ccccc@{}}
\toprule
Model                     & Type            & Recall@1                                               & Recall@5                                               & Recall@10                                              \\ \midrule
\multirow{2}{*}{GroupVit~\cite{xuGroupViTSemanticSegmentation2022}}    & Baseline        & 36.34                                                  & 65.35                                                  & 76.78                                                  \\
                          & Enhance w/ vote (\textbf{ours}) & \begin{tabular}[c]{@{}c@{}}\textbf{36.99}\\ ($\uparrow$0.65)\end{tabular} & \begin{tabular}[c]{@{}c@{}}\textbf{67.49}\\ ($\uparrow$2.14)\end{tabular} & \begin{tabular}[c]{@{}c@{}}\textbf{78.66}\\ ($\uparrow$1.88)\end{tabular} \\ \midrule
\multirow{2}{*}{CLIPSeg~\cite{lüddeckeImageSegmentationUsing2022}}  & Baseline        & 61.97                                                  & 86.18                                                 & 91.50                                                  \\
                          & Enhance w/ vote (\textbf{ours}) & \begin{tabular}[c]{@{}c@{}}\textbf{62.51}\\ ($\uparrow$0.54)\end{tabular} & \begin{tabular}[c]{@{}c@{}}\textbf{86.92}\\ ($\uparrow$0.74)\end{tabular} & \begin{tabular}[c]{@{}c@{}}\textbf{92.36}\\ ($\uparrow$0.86)\end{tabular} \\ \midrule
\multirow{2}{*}{ALGIN~\cite{jiaScalingVisualVisionLanguage2021d}}     & Baseline        & 73.91                                                  & 92.14                                                    & 95.70                                                  \\
                          & Enhance w/ vote (\textbf{ours}) & \begin{tabular}[c]{@{}c@{}}\textbf{74.11}\\ ($\uparrow$0.20)\end{tabular} & \begin{tabular}[c]{@{}c@{}}\textbf{92.60}\\ ($\uparrow$0.46)\end{tabular} & \begin{tabular}[c]{@{}c@{}}\textbf{96.10}\\ ($\uparrow$0.40)\end{tabular} \\ \midrule
\multirow{2}{*}{CLIP~\cite{radford2021learning}} & Baseline        & 58.65                                                  & 83.16                                                  & 89.76                                                  \\
                          & Enhance w/ vote (\textbf{ours}) & \begin{tabular}[c]{@{}c@{}}\textbf{58.67}\\ ($\uparrow$0.02)\end{tabular} & \begin{tabular}[c]{@{}c@{}}\textbf{83.74}\\ ($\uparrow$0.58)\end{tabular} & \begin{tabular}[c]{@{}c@{}}\textbf{90.28}\\ ($\uparrow$0.50)\end{tabular} \\ \bottomrule
\end{tabular}
\end{table}

\subsection{Case Study}
Fig.~\ref{fig:case-study} visually illustrates the retrieval process on Flickr30K-CFQ and public dataset Flickr30K , respectively, and uses heatmaps to demonstrate the method's efficacy intuitively. In the top left of Fig.~\subref{fig:cfq}, the original query (\textbf{\textcolor{green}{green}}, index $0$) and texts enhanced by the LLM (indexes $1-10$) are displayed. As shown in the top right of Fig.~\subref{fig:cfq}, we can find that the similarity matrix between these 11 texts and some images, with lighter colors indicating higher similarity. As highlighted by \textbf{\textcolor{red}{red}} frames, the similarities between the original query and corresponding images \textit{D} and \textit{F} are low but significantly increase for the enhanced texts. The second row of Fig.~\subref{fig:cfq} is the retrieved images, with the correct ones highlighted in red frame. Fig.~\subref{fig:flickr30k} provides similar evidence. Additionally, for the original query ``A group of people standing on the lawn in front of a building" in Fig.~\subref{fig:flickr30k}, image \textit{A} also matches this description, indicating annotation errors in the Flickr30K. In constructing our dataset, we merged some texts and corresponding images based on similarity, which partially mitigates this issue.

\section{Conclusion}
In this paper, we consider the textual insufficient of current text-image retrieval datasets in diversity and naturalness and introduce a new challenge set named \textbf{Flickr30K-CFQ}. It contains an additional four kinds of query corpus with multi-level granularities and oral description. The unsatisfactory performance performed by the existing method makes us propose a query-enhanced method using LLM to improve this real-world text-image retrieval task. Experimental results indicate that our proposed query-enhanced method achieves over $\mathbf{2\%}$ averagely improvement compared to existing methods tested on our proposed challenge set Flickr30K-CFQ. They also reflect the necessity of our Flickr30K-CFQ to provide a more efficient evaluation compared to the conventional general language-vision datasets for text-image retrieval community. 

\medskip
\noindent\textbf{Acknowledgement.}\quad This work is sponsored by the National Natural Science Foundation of China (No.U21A20488, 62072323), Zhejiang Lab Open Research Project (No.K2022NB0AB04), Shanghai Science and Technology Innovation Action Plan (No.22511104700) and Postdoctoral Fellowship Program of CPSF (GZC20232292).

%
%
\bibliographystyle{splncs04}
\bibliography{mybibliography}

\end{document}